# Resonant Spin Wave Excitation by Terahertz Magnetic Near-field Enhanced with Split Ring Resonator

Y. Mukai[1], H. Hirori[2,3,*], T. Yamamoto[4], H. Kageyama[2,4], and K. Tanaka[1,2,3,†]

[1] *Department of Physics, Graduate School of Science, Kyoto University, Sakyo-ku, Kyoto 606-8502, Japan*

[2] *Institute for Integrated Cell-Material Sciences (WPI-iCeMS), Kyoto University, Kyoto 606-8501, Japan*

[3] *CREST, Japan Science and Technology Agency, 4-1-8, Kawaguchi, Saitama 332-0012, Japan*

[4] *Department of Energy and Hydrocarbon Chemistry, Graduate School of Engineering, Kyoto University, Nishikyo-ku, Kyoto 615-8510, Japan*

## Abstract

Excitation of antiferromagnetic spin waves in $HoFeO_3$ crystal combined with a split ring resonator (SRR) is studied using terahertz (THz) electromagnetic pulses. The magnetic field in the vicinity of the SRR induced by the incident THz electric field component excites and the Faraday rotation of the polarization of a near-infrared probe pulse directly measures oscillations that correspond to the antiferromagnetic spin resonance mode. The good agreement of the temperature-dependent magnetization dynamics with the calculation using the two-lattice Landau-Lifshitz-Gilbert equation confirms that the spin wave is resonantly excited by the THz magnetic near-field enhanced at the LC resonance frequency of the SRR, which is 20 times stronger than the incident magnetic field.

PACS: 75.78.JP, 76.50.+g, 78.47.-p, 78.67.Pt

Ultrafast control of spins in solids has attracted considerable attention from researchers because of its importance in fundamental physics and for technological applications such as spintronics and spin-based information processing [1,2]. An optical technique with a high-intensity femtosecond pulse laser is a promising spin control method [3-6] and it has been used to reveal ultrafast magnetization dynamics in various magnetically ordered materials, including magnetic semiconductors and strongly correlated electron systems [7-10]. Recently, terahertz (THz) pulse generation and detection techniques have emerged as alternatives to optical techniques for studying magnetic excitations (spin waves) [11-13]. The magnetic field component has been shown to directly excite spin waves irrespective of the magneto-optical susceptibility of magnetic materials and without unavoidable excessive thermal energy due to electronic excitations that may arise from strong optical pulse excitations.

A split ring resonator (SRR), essentially subwavelength LC circuit, excited by a THz electric field induces a circulating current that in turn generates a magnetic field [14,15]. The properties of the SRR fabricated from high-conductivity metals, i.e., resonant field enhancement and subwavelength field localization, provide a good platform for studying magnetic excitations in condensed matter and also a route to developing electronically controlled hybrid spintronics and dynamic magneto-optical devices such as isolators. The LC response of the SRR is determined by its loop inductance and gap capacitance and the resonant enhancement effect at LC resonance frequency may be used to generate magnetic elementary excitations efficiently. However, an understanding of the interaction of the magnetic excitations with the THz magnetic near-field has remained elusive since the resonant and local enhancement properties of the SRR lead to complex driving forces on the magnetic excitations.

In this letter, we report on the dynamics of antiferromagnetic spin waves in $HoFeO_3$

crystal excited by the THz magnetic near-field resonantly enhanced with an SRR and measured by means of Faraday rotation optical microscopy. The temperature-dependent resonance frequency of the spin wave [16] allows us to investigate the resonant enhancement effect of the SRR systematically. The good agreement of the observed Faraday rotation with calculation based on the Landau-Lifshitz-Gilbert (LLG) equation confirms that the spin wave is resonantly excited by the THz magnetic near-field enhanced at the LC resonance frequency of SRR, which is 20 times stronger than the magnetic field strength of the incident THz pulse. These results imply that this method can change the magnetization 500 times more efficiently than optical excitation using the inverse Faraday effect.

Figure 1 shows the experimental setup of Faraday rotation optical microscopy with a THz pump pulse excitation of a hybrid sample consisting of an SRR and $HoFeO_3$ crystal. An amplified Ti:sapphire laser (repetition rate 1 kHz, central wavelength 780 nm, pulse duration 100 fs, and 4 mJ/pulse) was used as the light source. THz pulses were generated by optical rectification of the femtosecond laser pulses in a $LiNbO_3$ crystal by using the tilted-pump-pulse-front scheme and focused on the sample by using an off-axis parabolic mirror with a 50-mm focal length [17-19]. At the sample, the THz spot diameter $a$ at half intensity was ~300 μm, and the incident pump pulse energy was typically $E$=1.0 μJ (incident fluence $I_{THz} = E/\pi a^2$~0.35 mJ/cm$^2$). The $HoFeO_3$ crystal was grown with the floating-zone method [20]. We deposited $SiO_2$ anti-reflection (AR) and high-reflection (HR) coatings for the 780 nm probe pulse on the c-cut surface of single $HoFeO_3$ crystal (50 μm thickness) before the fabricating the SRR; these coatings enabled the probe pulse to be efficiently transmitted and reflected at the surfaces. A planar array of gold SRRs with a thickness of 100 nm was fabricated on the HR coating side by photolithography.

As shown in the inset of Fig. 1, the incident THz electric field $E_{in,x}$ polarized parallel to the arm with the gap causes the LC resonance. At the LC resonance frequency around 0.5 THz, the surface current flowing circularly in the arm induces a magnetic near-field $H_{nr,z}$ perpendicular to the surface in the vicinity of the SRR, whereas the pure electric-dipole resonance (which would occur in the gap) is around 1 THz [14]. The magnetization change induced by the magnetic near-field is measured by means of time resolved Faraday rotation optical microscopy. Focused by an objective lens, the spot size of the probe optical pulse at HR coating is ~1.5 μm diameter, enabling us to measure local changes in the Faraday rotation. The white dashed circle shown in the inset indicates the position at which the probe measures the Faraday rotation. A balanced detector, combined with a half-wave (λ/2) wave plate and a Wollaston prism, measures this rotation.

Holmium orthoferrite, $HoFeO_3$, is an excellent candidate for studying the interaction of spin waves with a THz magnetic near-field resonantly enhanced with the SRR because the spin resonance frequency can be tuned in the THz frequency region by changing the crystal temperature [16]. Between the Néel temperature ($T_N$~640 K) and the spin reorientation transition temperature ($T_1$=58 K), two magnetizations $m_i$ (i=1,2) of different iron sublattices ($Fe^{3+}$) are almost antiferromagnetically aligned along x-axis with a slight canting owing to the Dzyaloshinskii field, forming a spontaneous magnetization $M_s$ along the z-axis. (The experimental coordinate system is such that the x, y, and z-axes are parallel to the crystallographic a, b, and c-axes). The spin structure corresponds to $\Gamma_4$ magnetic symmetry with the ferromagnetic vector $M=M_s=m_1+m_2$ along the z-axis and the antiferromagnetic vector $L=m_1-m_2$ along the x-axis [21].

The THz magnetic near-field of the SRR along the z-axis can drives quasiantiferromagnetic (AF) mode (see the left panel of Fig. 2(a)) and induces magnetic deviations ($\Delta M_z$, $\Delta L_x$, and $\Delta L_y$). Two of these deviations ($\Delta M_z$ and $\Delta L_x$) can change the anti-symmetric off-diagonal elements of the dielectric tensor $\varepsilon_{xy}^a(=-\varepsilon_{yx}^a)$, which cause the Faraday effect for the light propagating for the z-direction [22]. Since $\Delta M_z/|\boldsymbol{M}_s| \gg \Delta L_x/L_0$, where $L_0$ is the magnitude of the antiferromagnetic vector $\boldsymbol{L}$ without external field, $\varepsilon_{xy}^a$ can be written as a function of $\Delta M_z/|\boldsymbol{M}_s|$,

$$\varepsilon_{xy}^a = ig_0\left(1+\zeta\frac{\Delta M_z}{|\boldsymbol{M}_s|}\right), \tag{1}$$

where $g_0$ is the strength of the Faraday effect without external fields, and $\zeta$ is the coefficient of ferromagnetic contribution to the Faraday effect [23]. Though the THz magnetic near-field in the xy-plane excites the quasiferromagnetic (F) mode (the right panel of Fig. 2(a)), the induced magnetic deviations ($\Delta M_x$, $\Delta M_y$, and $\Delta L_z$) do not change $\varepsilon_{xy}^a$. Thus, the probe light propagating along the z-axis in our experimental setup can detect only the AF-mode oscillation in the Faraday rotation.

Figure 2(b) shows the experimental Faraday rotation changes $\Delta\theta$ (open circles) as a function of delay time at different temperatures. The upper panel of Fig. 2(b) shows that the signal at 120 K oscillates harmonically with a period of 2 ps and a rise time of 15 ps, and it decays exponentially with a time constant of 70 ps. The temperature dependence of the temporal profile in Fig. 2(b) shows that the time period becomes shorter as the temperature increases, and the signal measured at 120 K has the largest maximum peak amplitude. Figure 2(c) shows the temperature dependence of the peak frequency obtained by taking a Fourier transform of the observed transient. The center frequency $\nu$ increases from 0.30 to 0.57 THz as the temperature increases from 60 to 200 K. This behavior is very similar to that of the absorption peak of AF-mode measured in the literature [16]. Remarkably, we find that the rise time (15 ps) is much longer than the temporal width of the incident THz pulse (~1 ps; see the inset of Fig. 2(b)). In addition, the temperature dependence shows that the Faraday rotation has the maximum value around 120 K (Fig. 2(b)) and its frequency is around 0.5 THz which is similar to the LC resonance frequency (Fig. 2(c)). These results imply that the AF-mode is excited by the THz magnetic near-field resonantly enhanced with the SRR.

To elucidate the dynamics of the magnetization driven by the THz magnetic near-field of the SRR, we calculated the magnetic near-field with a finite-difference time-domain (FDTD) method incorporating THz electric field transients measured by electro-optical (EO) sampling as an incident THz pulse source (see the inset of Fig. 2(b)) [24]. Figure 3(a) shows the spatial distribution of the enhancement factor, i.e., the ratio between the magnetic near-field $H_{nr,z}$ and the magnetic field of the incident THz pulse $H_{in,y}$ at the LC resonance frequency ($\nu_{LC}$=0.5 THz). More than 40-fold enhancement in the magnetic field amplitude occurs near the metallic arm, and it decreases farther from the arm. Figure 3(b) shows the spatial distribution of the enhancement factor along the x and z-directions through the circle indicated in Fig. 3(a), which corresponds to the position where the Faraday measurements are taken. Figures 3(c) and (d) show the temporal magnetic waveforms of the incident THz pulse and the generated THz near-field and their spectra. The maximum peak amplitude is twice that of the incident THz pulse (Fig. 3(c)) in the time domain and 24 times that in the frequency domain at the LC resonance frequency ($\nu_{LC}$=0.5 THz) (Fig. 3(d)) [25]. As shown in Fig. 3(c), the THz magnetic near-field continues to oscillate after the driving field has decayed away with a time constant (~15 ps) determined by the resonator's quality factor.

A theoretical analysis of the magnetization change $\Delta M_z/|M_s|$ and the resultant Faraday rotation enables us to quantitatively compare the experimentally observed temperature-dependent $\Delta\theta$ with the calculations. The temporal evolution of the sublattice magnetizations $m_i$ (i=1,2) in Fig. 2(a) under a magnetic driving field $H(t)$ are described by the LLG equation on the basis of the two-lattice model [26,27]: the spins at two different sites are coupled by the exchange interaction composing the free energy of the iron spin system $F$ [28]. Introducing the unit directional vector of the sublattice magnetizations, $R_i=m_i/m_0$ ($m_0=|m_i|$), the dynamics of the sublattice magnetizations is described by:

$$\frac{dR_i}{dt} = -\frac{\gamma}{(1+\alpha^2)}\Big(R_i\times[H(t)+H_{\mathrm{eff},i}] - \alpha R_i\times\big(R_i\times[H(t)+H_{\mathrm{eff},i}]\big)\Big), \qquad (2)$$

where $\gamma=2.213\times10^5$ m/As is the gyromagnetic ratio and $\alpha=1.8\times10^{-4}$ is the Gilbert damping constant. Using the normalized free energy $V=F/m_0$, the effective magnetic field is given by $H_{\mathrm{eff},i}=-\partial V/\partial R_i$ (i=1,2). Here, the calculated THz magnetic near-field (Fig. 3(c)) is used as the driving magnetic field $H(t)$ in Eq. (2). The Faraday rotation change $\Delta\theta$ caused by the magnetization change $\Delta M_z/|M_s|$ can be calculated by taking into account the birefringence of the samples and the spatially varying magnetization along the z-axis due to the inhomogeneous magnetic field distribution shown in Fig. 3 [29].

By making a least-squares fitting using a parameter $\zeta$ common to the data curves at different temperatures, theoretical curves (solid lines) are obtained that are very similar to the experimental data (open circles) in Fig. 3(b). The fitted value of $\zeta=0.29$ is reasonable when compared to the literature value of YFeO$_3$, $\zeta=0.23\pm0.13$ [30]. (The uncertainties are typically ±30%, owing mainly to errors in the matching of the spatial positions of the experiment and calculation.) $\zeta$ does not depend on the species of rare earth elements significantly because of their negligible effect on the magnetic order above the Néel temperature of rare earth spin system ~5K [31]. These results show that the spin waves are excited by the THz magnetic near-field of the SRR.

To examine the enhancement of the magnetization excitations, we plotted the area intensity of the Fourier spectrum of the observed Faraday transient as a function of their center frequencies (red open circles in Fig. 4). Figure 4 shows that the area intensity is enhanced around 0.5 THz, which is consistent with the enhancement factor calculated from the FDTD simulation. A quantitative comparison of the experimental results and the calculations (Fig. 2(b)) reveals that the enhanced magnetic field induces a magnetization change $\Delta M_z/|M_s|$ of 0.24 at 120 K. Rare-earth orthoferrites of $HoFeO_3$ are known to possess large inverse Faraday effects (spin-orbit coupling). Nonetheless, the maximum change of $\Delta M_z/|M_s|$ induced by an effective magnetic field pulse generated by using circularly polarized optical pulse (an amplitude of 0.6 T, a full-width at half maximum of 100 fs, a fluence of $I_{opt}$=10 mJ/cm$^2$) [32] is estimated to be only 0.012. These results mean that the THz-pulse method is 570 times more efficient than a circularly polarized optical pulse at changing the magnetization [33]. Here, the efficiency is defined by the ratio of the fluence of the incident THz (or optical) pulse with the magnetization change $\Delta M_z/|M_s|$.

In summary, we studied the excitation of antiferromagnetic spin waves in $HoFeO_3$ crystal combined with the SRR by using THz pulses. The magnetization dynamics observed by time-resolved Faraday rotation microscopy can be explained by the two-lattice LLG model with the magnetic near-field as the driving force. These results show that the THz magnetic near-field is resonantly enhanced by the SRR by ~20 times in amplitude compared with the incident THz magnetic field. This method can induce a magnetization change 500 times more efficiently than an optical excitation utilizing the inverse Faraday effect, and this indicates that it's a potential method to elucidate nonlinear magnetization responses without unavoidable excessive thermal energy due to electronic excitations that would accompany strong optical pulse excitations.

**Acknowledgements**

We are grateful to Masashi Kawaguchi and Teruo Ono for letting us use an evaporation apparatus and their help. This study was supported by KAKENHI (24760042 and 20104007) from JSPS and MEXT of Japan, and Industry-Academia Collaborative R&D from Japan Science and Technology Agency (JST). H.H. is grateful to Rupert Huber for fruitful discussions during his visit in Regensburg through the iCeMS-JSPS Overseas Visit Program for Young Researchers (Bon Voyage Program).

**Figure Captions**

**Fig. 1 (Color online)** Schematic setup of THz pump-NIR Faraday rotation measurement. The inset shows the SRRs fabricated on the c-cut surface of the HoFeO$_3$ crystal. The length of the ring $l_m$ is 34 μm, the width of metal lines $d_m$ is 6 μm, and the split gap $g_m$ is 6 μm. The dashed circle indicates the probe position where the Faraday rotation is measured.

**Fig. 2 (Color online)** (a) Magnetization motions for two modes are shown using the ferromagnetic vector $M=m_1+m_2$ and the antiferromagnetic vector $L=m_1-m_2$. (b) Upper panel: temporal change of the Faraday rotation angle measured at 120 K (open circles) and calculation (black line). The inset shows the electric field of the THz transient measured by EO sampling. Lower panel: the expanded Faraday rotation signals of experiment (open circles) and calculation (black line) between 0–20 ps at different temperatures (200, 120, and 100 K). (c) Temperature dependence of the peak frequency obtained by taking a Fourier transform of the observed Faraday rotation transient.

**Fig. 3 (Color online)** (a) Finite-difference time-domain method (FDTD) calculation of the magnetic field near the SRR. Two-dimensional spatial distribution of z-directed magnetic field $H_{nr,z}$ at the LC resonance frequency ($\nu_{LC}$=0.5 THz) at the interface between the HR coat and HoFeO$_3$ (z=1.2 μm), plotted as the enhancement factor, i.e., the ratio between the magnetic near-field $H_{nr,z}(\nu_{LC})$ and the magnetic field of the incident THz pulse $H_{in,y}(\nu_{LC})$. The dashed circle indicates the probe position where the Faraday rotation is measured. (b) The spatial distribution of the enhancement factor $H_{nr,z}(\nu_{LC})/H_{in,y}(\nu_{LC})$ along the x and z-directions through the circle indicated in (a). (c) The temporal magnetic waveforms of the incident THz pulse and the generated THz near-field at the probe position (z=1.2μm) and (d) their spectra.

**Fig. 4 (Color online)** The area intensity of the Fourier spectrum of the observed Faraday transient as a function of their center frequencies (red open circles). The spectrum of the calculated enhancement factor, i.e., the ratio between the magnetic near-field $H_{nr,z}(\nu)$ and the magnetic field of the incident THz pulse $H_{in,y}(\nu)$ (solid line) at the fixed position indicated by the dashed circle in Fig. 3(a) ($z=1.2\mu m$).

# Fig.1

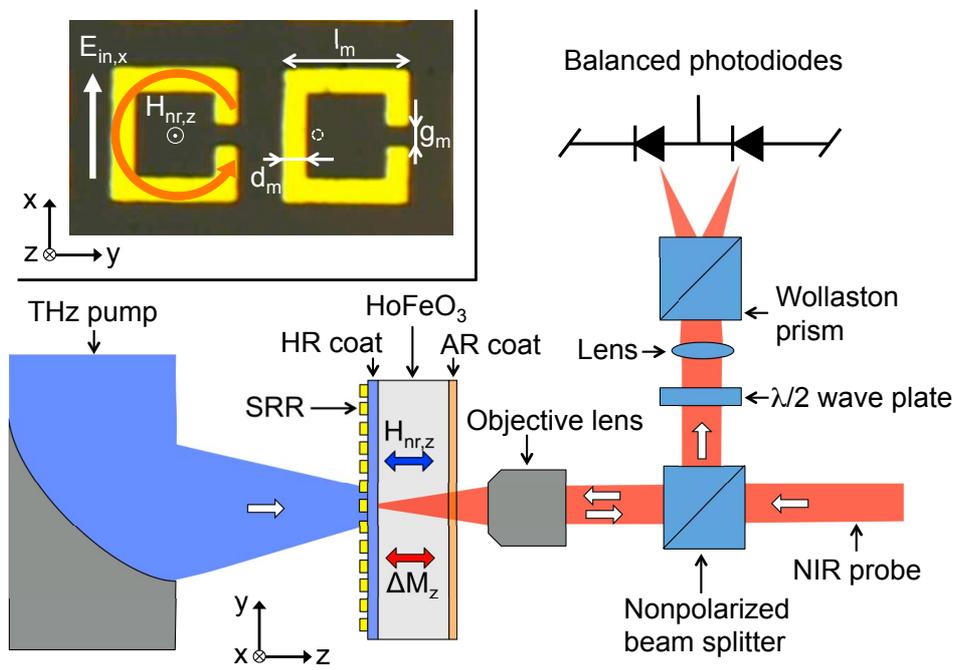

Y. Mukai

**Fig. 2**

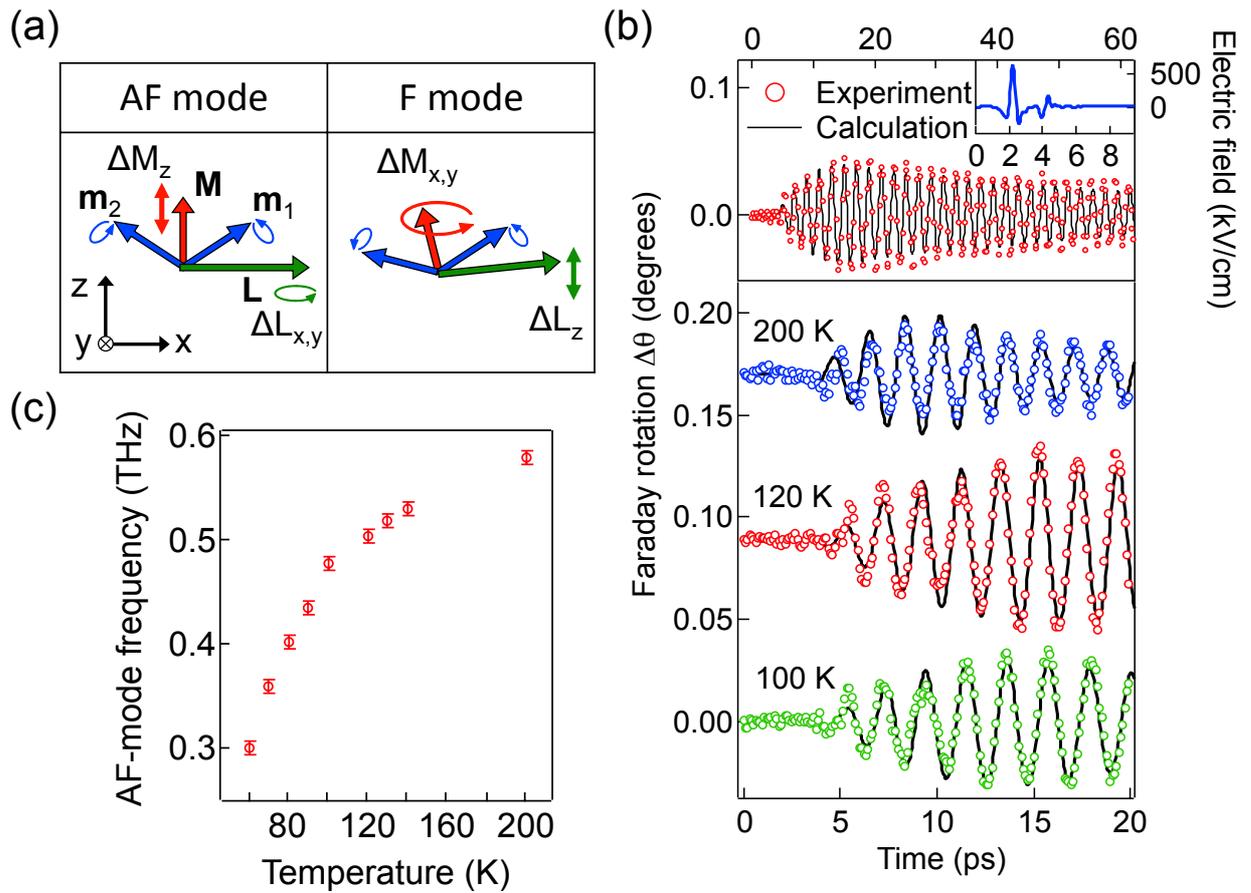

Y. Mukai

**Fig. 3**

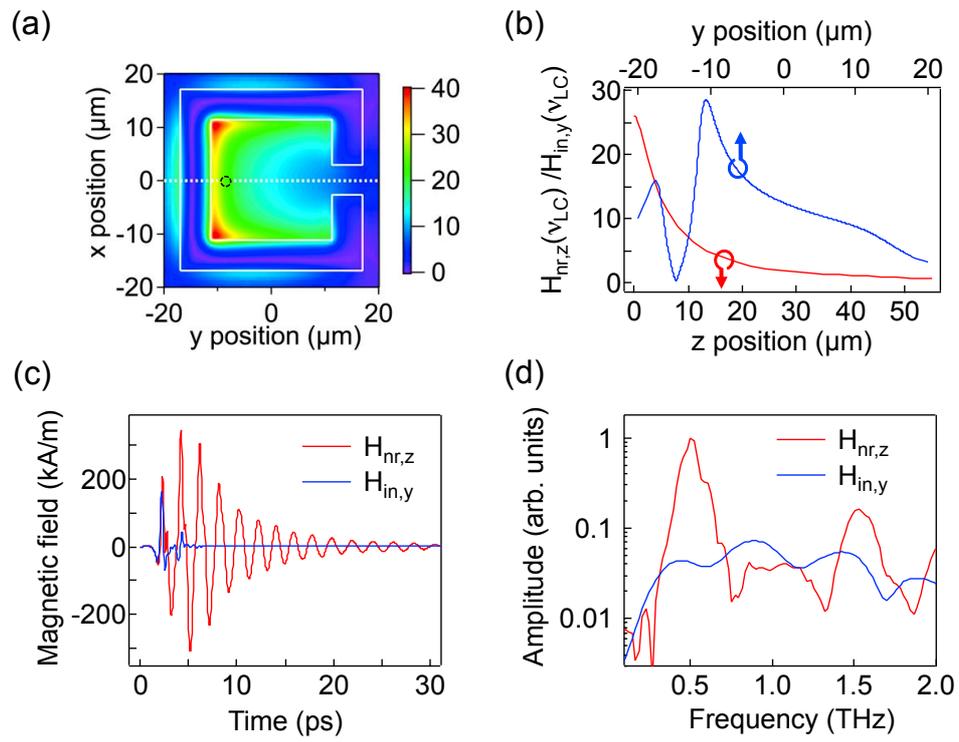

Y. Mukai

**Fig. 4**

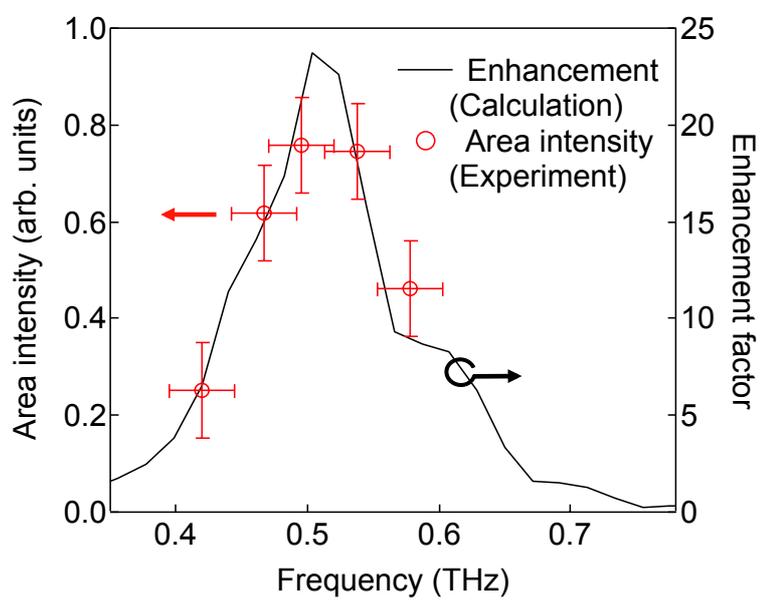

Y. Mukai

# Resonant Spin Wave Excitation by Terahertz Magnetic Near-field Enhanced with Split Ring Resonator

## —Supplemental material—

### I. Free energy of iron ion system in HoFeO$_3$

The free energy $F$ of the iron ion ($Fe^{3+}$) system based on the two-lattice model is a function of two different iron sublattice magnetizations $m_i$, and composed of the exchange energy and one-site anisotropy energy [1,2]. The free energy normalized by the sublattice magnetization magnitude, $V=F/m_0$ ($m_0=|m_i|$), is expanded as a power series in the unit directional vector of the sublattice magnetizations, $R_i=m_i/m_0=(X_i,Y_i,Z_i)$. In the magnetic phase $\Gamma_4$ ($T>58K$), the normalized free energy is given as follows [1,2]:

$$V = E R_1 \cdot R_2 + D(X_1 Z_2 - X_2 Z_1) - A_{xx}(X_1^2 + X_2^2) - A_{zz}(Z_1^2 + Z_2^2). \quad (S1)$$

The first two terms mean the exchange energy, where $E(=5.1\times10^8$ A/m) and $D(=1.1\times10^7$ A/m) for HoFeO$_3$ are the symmetric and antisymmetric exchange field, respectively [3]. The anisotropy constants $A_{xx}$ and $A_{zz}$ are related to the F and AF-mode frequencies ($\omega_F$ and $\omega_{AF}$) and are given by:

$$\frac{\omega_{AF}^2}{\gamma^2} = 4EA_{xx} + 4A_{xx}(A_{xx} - A_{zz}) + D^2, \quad (S2)$$

$$\frac{\omega_F^2}{\gamma^2} = 4E(A_{xx} - A_{zz}) + 4A_{xx}(A_{xx} - A_{zz}). \quad (S3)$$

$A_{xx}$ and $A_{zz}$ at each measurement temperature can be estimated by using the temperature dependent AF-mode frequency $\omega_{AF}$ measured experimentally in Fig. 2(c) and a constant F-mode frequency $\omega_F/2\pi$ (=0.37 THz) from the literature [4]. By numerically solving Eq. (2) in the main text with the free energy described by Eqs. (S1)-(S3), the macroscopic magnetization change $\Delta M(z,t)/|M_s|$ can be calculated as a function of time $t$ and longitudinal position $z$.

## II. Relation between magnetization change and Faraday rotation

The polarization analysis of a probe pulse propagating along the z-axis allows us to calculate a Faraday rotation change that can be compared with the experimental data in Fig. 2(b). The dielectric constant of the $HoFeO_3$ crystal including the Faraday effect and a crystal birefringence is descried by the following dielectric tensor [5]:

$$\tilde{\varepsilon}(z,t)=\begin{pmatrix} \varepsilon_{xx} & \varepsilon^a_{xy}(z,t) \\ \varepsilon^a_{yx}(z,t) & \varepsilon_{yy} \end{pmatrix} = \begin{pmatrix} \varepsilon+\eta & ig(z,t) \\ -ig(z,t) & \varepsilon-\eta \end{pmatrix}, \qquad (S4)$$

where $\eta$ and $g(z,t)$ respectively represent the strength of birefringence and the Faraday effect. In the $\Gamma_4$ phase, $g(z,t)$ is mainly modulated by the THz-induced magnetization change of $\Delta M_z(z,t)/|M_s|$, since $\Delta M_z/|M_s| \gg \Delta L_x/L_0$. Here, $L_0$ is the magnitude of the antiferromagnetic vector $L$ without external fields. Thus, the off-diagonal elements of the dielectric tensor are rewritten as [6]:

$$\varepsilon^a_{xy}(z,t)=ig(z,t)=ig_0\left(1+\zeta\frac{\Delta M_z(z,t)}{|M_s|}\right), \qquad (S5)$$

where $\zeta$ is the coefficient of the ferromagnetic contribution to the Faraday effect and $g_0$ is $g(z,t)$ without external fields.

The electric field of a probe pulse propagating along the z direction in a medium with a dielectric tensor described by Eq. (S5) obeys the wave equation:

$$\tilde{\varepsilon}(z,t)\mu \frac{\partial^2 \boldsymbol{E}}{\partial t^2}=\frac{\partial^2 \boldsymbol{E}}{\partial z^2}. \quad (S6)$$

In the analysis of the polarization state of a probe pulse passing through birefringent media of $HoFeO_3$, the electric field can be expressed in terms of the following two normal modes $\boldsymbol{n}_{\pm}$, which are elliptically polarized and propagate with different wavenumbers $k_{\pm}$ [6]:

$$\boldsymbol{n}_{\pm}(z,t)=\begin{pmatrix} E_x \\ E_y \end{pmatrix}=\begin{pmatrix} 1 \\ i\left(\Gamma(z,t)\mp\sqrt{1+\Gamma(z,t)^2}\right) \end{pmatrix} e^{i(\omega t-k_{\pm}(z,t)z)}, \quad (S7)$$

$$k_{\pm}(z,t)=\sqrt{\omega^2\mu\left(\varepsilon\pm\sqrt{\eta^2+g(z,t)^2}\right)}, \quad (S8)$$

$$\Gamma(z,t)=\eta/g(z,t), \quad (S9)$$

$$\delta(z,t)=k_+(z,t)-k_-(z,t), \quad (S10)$$

where $\mu$ is the magnetic permeability and $\omega$ is the center angular frequency of the probe pulse. The parameter $\Gamma(z,t)$ represents the ratio of strengths between birefringence and Faraday effects, and $\delta$ is the wavenumber mismatch. Since $\sqrt{\eta^2+g^2}\ll\varepsilon$, Eqs. (S9) and (S10) can be rewritten as follows:

$$\Gamma(z,t)=\Gamma_0\left(1+\zeta\frac{\Delta M_z(z,t)}{|\boldsymbol{M}_s|}\right)^{-1}, \quad (S11)$$

$$\delta(z,t)=\delta_0\sqrt{\frac{1+\Gamma(z,t)^{-2}}{1+\Gamma_0^{-2}}}, \quad (S12)$$

where $\Gamma_0=\eta/g_0\sim 32$ and $\delta_0=\omega\sqrt{\mu(g_0^2+\eta^2)/\varepsilon}\sim 1.5\times 10^5$ deg/cm for the wavelength $\lambda=780$ nm are used [5,7]. The Faraday rotation angle change $\Delta\theta$ of the outgoing light ($E_{x,out}$, $E_{y,out}$) receiving the retardation during the passing through the entire medium is

$$\Delta\theta = \frac{1}{2}\tan^{-1}\left(\frac{2\mathrm{Re}(E_{x,\mathrm{out}}^{*}E_{y,\mathrm{out}})}{|E_{x,\mathrm{out}}|^{2}-|E_{y,\mathrm{out}}|^{2}}\right) - \theta_{0}, \tag{S13}$$

where $\theta_0$=0.12 degree is the Faraday rotation angle without the THz excitation.